\documentclass[a4paper]{article}
\pdfoutput=1
\usepackage{INTERSPEECH2022}
\usepackage{todonotes}

\title{End-To-End Label Uncertainty Modeling for Speech-based Arousal Recognition Using Bayesian Neural Networks}

\name{Navin Raj Prabhu$^{\star \dagger}$ \quad Guillaume Carbajal$^{\star}$ \quad Nale Lehmann-Willenbrock$^{\dagger}$  \quad Timo Gerkmann$^{\star}$ }
  
\address{$^{\star}$Signal Processing, Universit\"at Hamburg, Germany \\
  $^{\dagger}$Industrial and Organizational Psychology, Universit\"at Hamburg, Germany}

\email{\{navin.raj.prabhu, guillaume.carbajal, nale.lehmann-willenbrock, timo.gerkmann\}@uni-hamburg.de}

\begin{document}

\maketitle
\begin{abstract}
  Emotions are subjective constructs. Recent end-to-end speech emotion recognition systems are typically agnostic to the subjective nature of emotions, despite their state-of-the-art performance. In this work, we introduce an end-to-end Bayesian neural network architecture to capture the inherent subjectivity in the arousal dimension of emotional expressions. To the best of our knowledge, this work is the first to use Bayesian neural networks for speech emotion recognition. At training, the network learns a distribution of weights to capture the inherent uncertainty related to subjective arousal annotations. To this end, we introduce a loss term that enables the model to be explicitly trained on a distribution of annotations, rather than training them exclusively on mean or gold-standard labels. We evaluate the proposed approach on the AVEC'16 dataset. Qualitative and quantitative analysis of the results reveals that the proposed model can aptly capture the distribution of subjective arousal annotations, with state-of-the-art results in mean and standard deviation estimations for uncertainty modeling.
\end{abstract}


\section{Introduction}
\label{sec:intro}

While individual subjective emotional experiences may be accessed using self-report surveys \cite{ledoux2018subjective}, expressions of emotions are embedded in a social context, which makes them inherently dynamic and subjective interpersonal phenomena \cite{lei2015affect, nummenmaa2018maps}. One way in which emotions become expressed in social interactions and therefore accessible for social signal processing concerns speech signals. Speech emotion recognition (SER) research spans roughly two decades \cite{Schuller2018-xi}, with ever improving state-of-the-art results. As a consequence, affective sciences and SER has shown increasing prominence in high-critical and socially relevant domains, e.g. health and employee well-being \cite{Schuller2018-xi, dukes2021rise}. 

Common SER approaches rely on hand-crafted features to predict gold-standard emotion labels \cite{han2020exploring, sridhar2020modeling}. Recently, end-to-end deep neural networks (DNNs) have been shown to deliver state-of-the-art emotion predictions \cite{Tzirakis2018-speech, tzirakis2021-mm}, by \emph{learning} features rather than relying on hand-crafted features. Despite their state-of-the-art results, they are often agnostic to the subjective nature of emotions and the resulting label uncertainty \cite{han2017hard, sridhar2020modeling}, thereby inducing limited reliability in SER \cite{Schuller2018-xi}. However, for any real-world application context, it is crucial that SER systems should not only deliver mean or gold-standard predictions but also account for subjectivity based confidence measures \cite{gunes2013categorical, Schuller2018-xi}.




Han et al., \cite{han2017hard, han2020exploring} pioneered uncertainty modeling in SER using a multi-task framework to also predict the standard deviation of emotion annotations. Sridhar et al., \cite{sridhar2020modeling} introduced a dropout-based model to estimate uncertainties. However, these uncertainty models were not trained on the distribution of emotion annotations and relied on hand-crafted features. Of note, in contrast to hand-crafted features, learning representations in an end-to-end manner implies that they are also dependent on the level of subjectivity in label annotations \cite{alisamir2021evolution}.


In machine learning, two types of uncertainty can be distinguished. \textit{Aleatoric} uncertainty captures data inherent noise (label uncertainty) whereas \textit{epistemic} uncertainty accounts for the model parameters and structure (model uncertainty) \cite{zheng2021uncertainty}. Stochastic and probabilistic models have mainly been deployed for uncertainty modeling, using auto-encoder architectures \cite{kohl2018probabilistic}, neural processes \cite{garnelo2018conditional}, and Bayesian neural networks (BNN) \cite{blundell2015weight}. Bayes by Backpropagation (BBB) for BNNs \cite{blundell2015weight} uses simple gradient updates to optimize weight distributions. Further with its capability to produce stochastic outputs, it is a promising candidate for end-to-end uncertainty SER. 
%

As opposed to emotions as inner subjective experiences \cite{ledoux2018subjective}, we focus on emotional expressions as behaviors that others subjectively perceive and respond to. A common framework for analyzing the expression of emotion is pleasure-arousal theory \cite{reisenzein1994pleasure, russell1980circumplex}, which describes emotional experiences in two continuous, bipolar, and orthogonal dimensions: pleasure-displeasure (\emph{valence}) and activation-deactivation (\emph{arousal}). It is documented in SER literature that the audio modality insufficiently explains valence \cite{Tzirakis2018-speech, tzirakis2021-mm}. Noting this, in this work, we decided to specifically focus on the label uncertainty in the \emph{arousal} dimension of emotional expressions.






In this paper, we propose an end-to-end BBB-based BNN architecture for SER. To the best of our knowledge, this is the first time a BNN is used for SER. In contrast to \cite{han2017hard, han2020exploring, sridhar2020modeling}, the model can be explicitly trained on a distribution of emotion annotations. For this, we introduce a loss term that promotes capturing \emph{aleotoric} uncertainty (label uncertainty) rather than exclusively capturing \emph{epistemic} uncertainty (model uncertainty). Finally, we show that our proposed model trained on the loss term can aptly capture label uncertainty in arousal annotations.

The rest of the paper is organized as follows. In Section \ref{sec:lit}, we present related background on label uncertainty. In Section \ref{sec:method}, we introduce the proposed end-to-end BNN SER model. In Section \ref{sec:experiments}, we explain the experimental setup. In Section \ref{sec:Discussion}, we present the results and raise discussions on them.

\section{On uncertainty in arousal annotations}
\label{sec:lit}





\subsection{Ground-truth labels}
\label{sec:gt-ser-lit}

A crucial challenge in studying emotions within the arousal and valence framework concerns the significant degree of subjectivity surrounding them \cite{Schuller2018-xi, han2017hard, t21_interspeech}. To tackle this, annotations $\{y_{1}, y_{2}, .., y_{a}\}$ for emotions are collected from $a$ annotators \cite{recolaDB, raj2020defining}. The \emph{ground-truth label} is then obtained as the mean $m$ over all annotations from $a$ annotators \cite{abdelwahab2019active, avec16},
\begin{equation} \label{eq:mean-annot}
    {m} = \dfrac{1}{a} \sum_{i=1}^a y_{i}.
\end{equation}
Alternatively, an evaluator-weighted mean has been proposed and referred to as the \emph{gold-standard} $\widetilde{m}$ \cite{grimm2005evaluation, Tzirakis2018-speech}. However, in order to better represent subjective annotations, in this paper we use the mean $m$ rather than the evaluator-weighted mean.



\subsection{Point estimates in SER}
Given a raw audio sequence of $T$ frames $\mathcal{X} = [x_1, x_2, ..., x_T]$, typically, the goal is to estimate the ground-truth label $m_t$ for each time frame $t \in [1,T]$, referred to as $\widehat{m}_t$. The concordance correlation coefficient (CCC), which takes both linear correlations and biases into consideration, has been widely used as a loss function for this task. For Pearson correlation $r$, the CCC between the ground-truth label $m$ and its estimate $\widehat{m}$, for $T$ frames, is formulated as
\begin{equation}\label{loss:CCC}
    \mathcal{L}_{\text{CCC}}(m) = {\frac {2r \sigma_{m}\sigma_{\widehat{m}}}{\sigma _{m}^{2}+\sigma_{\widehat{m}}^{2}+(\mu _{m}-\mu _{\widehat{m}})^{2}}},
\end{equation}
where $\mu_{m} = \dfrac{1}{T} \sum_{t=1}^{T}m_t$, $\sigma_{m}^{2} = \dfrac{1}{T} \sum_{t=1}^T (m_t - \mu_m)^2$, and $\mu _{\widehat{m}}$, $\sigma_{\widehat{m}}^{2}$ are obtained similarly for $\widehat{m}$.



Early approaches relied on hand-crafted features as inputs to estimate CCC. End-to-end DNNs which circumvent the limitations of hand-crafted and -chosen features have been deployed to yield state-of-the-art performance \cite{trigeorgis2016adieu, Tzirakis2018-speech, tzirakis2021-mm}. Notwithstanding their performance, end-to-end DNNs are trained exclusively on $m$ and therefore cannot account for annotator subjectivity.

\subsection{Uncertainty modeling in SER}
\label{sec:labelUncert-lit}



Han et al. quantified label uncertainty using a statistical estimate, the perception uncertainty $s$ defined as the unbiased standard deviation of $a$ annotators \cite{han2017hard, han2020exploring}:
\begin{equation}\label{eq:pu}
    s = \sqrt{\dfrac{1}{a - 1} \sum_{i=1}^a (y_i - m)^2}.
\end{equation}
They proposed a multi-task model to additionally estimate $s$ along with $m$. However, the model only accounts for the standard deviation of $a$ annotations, rather than the whole distribution in itself. Thereby, susceptible to unreliable $s$ estimates for lower values of $a$ and sparsely distributed annotations.





Sridhar et al. introduced a Monte-Carlo dropout-based uncertainty model, to obtain stochastic predictions and uncertainty estimates \cite{sridhar2020modeling}. The model is trained exclusively on $m$ rather than the distribution of annotations, thereby only capturing the model uncertainty and not the label uncertainties.

\section{End-to-end label uncertainty model}
\label{sec:method}

In order to better represent subjectivity in emotional expressions, we propose to estimate the \emph{emotion annotation distribution} $\mathcal{Y}_t$ for each frame $t$. While the true distributional family of $\mathcal{Y}_t$ is unknown, we assume, for simplicity, that it follows a Gaussian distribution:
\begin{equation}
\mathcal{Y}_t \sim \mathcal{N}(m_t, s_t).
\end{equation}
Thus, the goal is to obtain an estimate $\widehat{\mathcal{Y}}_t$ of $\mathcal{Y}_t$ and infer both $\widehat{m}_t$ and $\widehat{s}_t$ from realizations of $\widehat{\mathcal{Y}}_t$.






\subsection{End-to-end DNN architecture}
\begin{figure}[t!]
    \centering
    \includegraphics[width=0.48\textwidth]{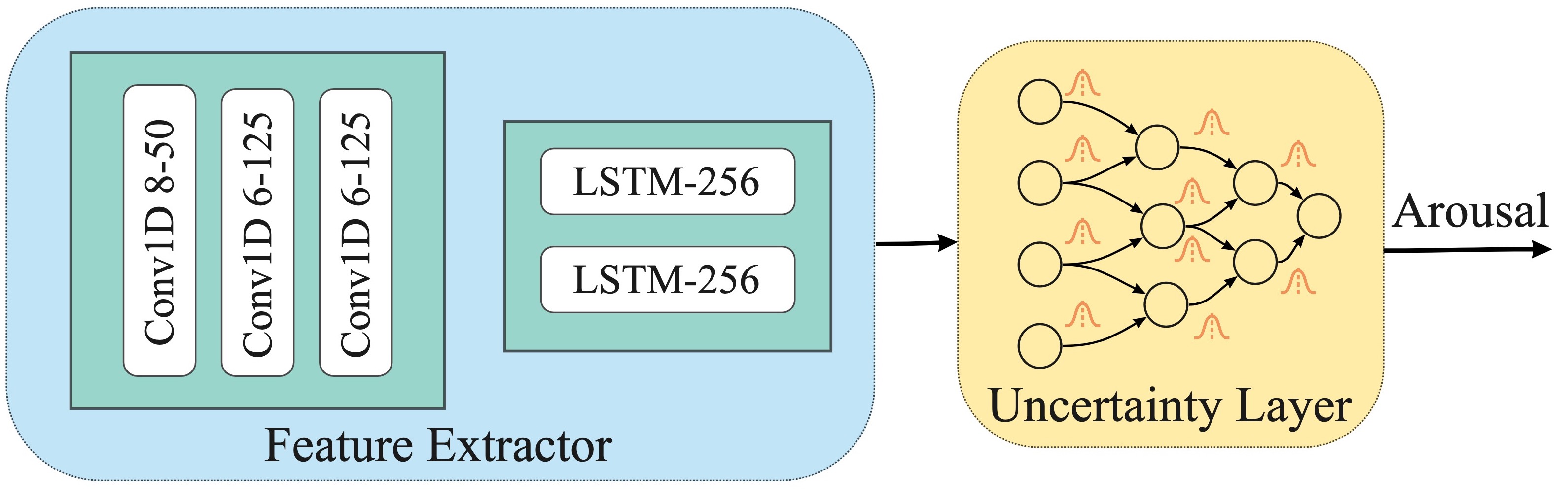}
    \caption{The proposed architecture. Blue denotes layers with deterministic weights and yellow with probabilistic weights.}
    \label{Fig:speechEmoBnn}
\end{figure}

We propose an end-to-end architecture which uses a feature extractor to learn temporal-paralinguistic features from $x_t$, and an uncertainty layer to estimate $\mathcal{Y}_t$ (see Fig. \ref{Fig:speechEmoBnn}). The feature extractor, inspired from \cite{Tzirakis2018-speech}, consists of three Conv1D layers followed by two stacked long-short term memory (LSTM) layers. The uncertainty layer is devised using the BBB technique \cite{blundell2015weight}, comprising three BBB-based MLP.


\subsection{Model uncertainty loss}
Unlike a standard neuron which optimizes a deterministic weight $w$, the BBB-based neuron learns a probability distribution on the weight $w$ by calculating the variational posterior $P(w|\mathcal{D})$ given the training data $\mathcal{D}$ \cite{blundell2015weight}. Intuitively, this regularizes $w$ to also capture the inherent uncertainty in $\mathcal{D}$. 

Following \cite{blundell2015weight}, we parametrize $P(w|\mathcal{D})$ using a Gaussian $\mathcal{N}(\mu_w, \sigma_w)$. For non-negative $\sigma_w$, we re-parametrize the standard deviation $\sigma_w = \log(1 + \exp(\rho_w))$.
Then, $\theta = (\mu_w, \rho_w)$ can be optimized using simple backpropagation.

For an optimized $\theta$, the predictive distribution $\widehat{\mathcal{Y}}_t$ for an audio frame $x_t$, is given by $P (\widehat{y}_t|x_t) = \mathbb{E}_{P(w|D)}[P(\widehat{y}_t|x_t, w)]$, where $\widehat{y}_t$ are realizations of $\widehat{\mathcal{Y}}_t$. Unfortunately, the expectation under the posterior of weights is intractable. To tackle this, the authors in \cite{blundell2015weight} proposed to learn $\theta$ of a weight distribution $q(w|\theta)$, the variational posterior, that minimizes the Kullback-Leibler (KL) divergence with the true Bayesian posterior, resulting in the negative evidence lower bound (ELBO),
\begin{equation} \label{loss:BBB}
    f(w,\theta)_{\text{BBB}} = \text{KL} \big[q(w|\theta) \| P(w)\big] - \mathbb{E}_{q(w|\theta)} \big[\log P(D|w)\big]. 
\end{equation}

Stochastic outputs in BBB are achieved using multiple forward passes $n$ with stochastically sampled weights $w$, thereby modeling $\widehat{\mathcal{Y}}_t$ using the $n$ stochastic estimates. To account for the stochastic outputs, \eqref{loss:BBB} is approximated as,
\begin{equation}\label{loss:BBB-stoch}
    \mathcal{L}_{\text{BBB}} \approx \sum_{i=1}^{\textit{$n$}} \log q(w^{(i)}|\theta) - \log P(w^{(i)}) - \log P(D|w^{(i)}).
\end{equation}
where $w^{(i)}$ denotes the $i^{th}$ weight sample drawn from $q(w|\theta)$. The BBB window-size $b$ controls how often new weights are sampled for time-continuous SER. The degree of uncertainty is assumed to be constant within this time period.

During testing, the uncertainty estimate $\widehat{s}_t$ is the standard deviation of $\widehat{\mathcal{Y}}_t$. Similarly, $\widehat{m}_t$ is the realization $\widehat{y}_t$ obtained using the mean of the optimized weights $\mu_w$. Obtaining $\widehat{m}_t$ using $\mu_w$ helps overcome the randomization effect of sampling from $q(w|\theta)$, which showed better performances in our case.

\subsection{Label uncertainty loss}
\label{sec:bnn_labelUncert}



Inspired by \cite{zheng2021uncertainty}, we introduce a KL divergence-based loss term as a measurement of distribution similarity to explicitly fit our model to the label distribution $\mathcal{Y}_t$:
\begin{equation}\label{loss:KL}
    \mathcal{L}_{\text{KL}} = f(\mathcal{Y}_t\|\widehat{\mathcal{Y}}_t)_{\text{KL}} = \int  \mathcal{Y}_t(x) \: \log\frac{\mathcal{Y}_t(x)}{\widehat{\mathcal{Y}}_t(x)} dx.
\end{equation}
The KL divergence is asymmetric, making the order of distributions crucial. In \ref{loss:KL}, the true distribution $\mathcal{Y}_t$ is followed by its estimate $\widehat{\mathcal{Y}}_t$, promoting a mean-seeking approximation rather than a mode-seeking one and capturing the full distribution \cite{Goodfellow-et-al-2016}.

\subsection{End-to-end uncertainty loss}
The proposed end-to-end uncertainty loss is formulated as,
\begin{equation}\label{eq:end-to-end_loss}
   \mathcal{L} = \mathcal{L}_{\text{CCC}}(m) + \mathcal{L}_{\text{BBB}} + \alpha \mathcal{L}_{\text{KL}}.
\end{equation}
 Intuitively, $\mathcal{L}_{\text{CCC}}(m)$ optimizes for mean predictions $m$, $\mathcal{L}_{\text{BBB}}$ optimizes for BBB weight distributions, and $\mathcal{L}_{\text{KL}}$ optimizes for the label distribution $\mathcal{Y}_t$.
 For $\alpha=0$, the model only captures model uncertainty (\emph{MU}). For $\alpha=1$, the model also captures \emph{label uncertainty} (\emph{$+$LU}).
 $\mathcal{L}_{\text{CCC}}(m)$ is used as part of $\mathcal{L}$ to achieve faster convergence and jointly optimize for mean predictions. 
 Including $\mathcal{L}_{\text{CCC}}(m)$
 might lead to better optimization of the feature extractor, as previously illustrated by 
 \cite{Tzirakis2018-speech, tzirakis2021-semspeech}.

\section{Experimental Setup}
\label{sec:experiments}

\subsection{Dataset}\label{sec:dataset}
To validate our model, we use the AVEC'16 emotion recognition dataset \cite{avec16}. Multimodal signals were recorded from 27 subjects, and in this work we only utilize the audio signals collected at 16 kHz. The dataset consists of continuous arousal annotations by $a=6$ annotators at $40$~ms frame-rate, and post-processed with local normalization and mean filtering. The arousal annotations are distributed on average with $\mu_{{m}} = 0.01$ and $\mu_{s} = 0.23$, where $\mu_{s} = \dfrac{1}{T} \sum_{t=1}^{T}s_t$. The dataset is divided into speaker disjoint partitions for training, development and testing, with nine $300$~s recordings each. As the annotations for the test partition in the dataset are not publicly available, all results are computed on the development partition. 

%

\subsection{Baselines}

As baselines, we use Han et al.'s perception uncertainty model (\emph{MTL PU}) and single-task learning model (\emph{STL}) \cite{han2020exploring}. The \emph{MTL PU} model uses a multi-task technique followed by a dynamic tuning layer to account for perception uncertainty $s$ in the final mean estimations $m$. The \emph{STL} on the other hand only performs a single task of estimating $m$. 

For a fair comparison, we reimplemented the baselines and tested them in our test bed. Crucially, the reimplementation also enables us to compare the models in-terms of their $s$ estimates, which were not presented in Han et al.'s work \cite{han2020exploring}. The only difference between Han et al.'s test bed and ours is the post-processing pipeline. While Han et al. use the post-processing pipeline suggested in the AVEC'16 \cite{avec16}, here we use the median filtering \cite{Tzirakis2018-speech} as the sole post-processing technique to make all considered approaches comparable.

\subsection{Choice of hyperparameters}
\label{sec:bnn_hyperparam}
The hyperparameters of the \emph{feature extractor} (e.g. kernel sizes, filters) are adopted from \cite{tzirakis2021-semspeech}. A similar extractor with the same hyperparameters has been used in several multimodal emotion recognition tasks with state-of-the-art performance \cite{tzirakis2021-mm, tzirakis2021-semspeech}. 

As the \emph{prior distribution} $P(w)$, \cite{blundell2015weight} recommend a mixture of two Gaussians, with zero means and standard deviations as $\sigma_1 > \sigma_2$ and $\sigma_2 \ll 1$, thereby obtaining a spike-and-slab prior with heavy tail and concentration around zero mean. But in our case, we do not need mean centered predictions as $\mathcal{Y}$ does not follow such a distribution, as seen in Section \ref{sec:dataset}. In this light, we propose to use a simple Gaussian prior with unit standard deviation $\mathcal{N}(0, 1)$. The $\mu_w$ and $\rho_w$ of the \emph{posterior distribution} $P(w|D)$ are initialized uniformly in the range $[-0.1, 0.1]$ and $[-3, -2]$ respectively. The ranges were fine-tuned using grid search for maximized $\mathcal{L}_\text{KL}$  at initialization on the train partition. 




It is computationally expensive to sample new weights at every time-step ($40$~ms) and also the level of uncertainties varies rather slowly. In this light, we set \emph{BBB window-size} $b = 2$~s ($50$ frames). The same window-size is also used for median filtering, the sole post-processing technique used. The \emph{number of forward passes} in \eqref{loss:BBB-stoch} is fixed to $n = 30$, with the time-complexity in consideration.

%
For training, we use the Adam optimizer with a learning rate of $10^{-4}$. The batch size used was 5, with a sequence length of 300 frames, $40$~ms each and $12$~s in total. Dropout with probability $0.5$ was used to prevent overfitting. All the models were trained for a fixed 100 epochs. The best model is selected and used for testing when best $\mathcal{L}$ is observed on train partition.

\begin{table}[t!]
    \caption{Comparison on mean $m$, standard deviation $s$, and label distribution estimations $\mathcal{Y}$, in terms of $\mathcal{L}_{\text{ccc}}(m)$, $\mathcal{L}_{\text{ccc}}(s)$, and $\mathcal{L}_\text{KL}$ respectively. Larger CCC indicates improved performance. Lower KL indicates improved performance. {**} indicates statistically significant better results over \textbf{all} other approaches in comparison, and {*} over \textbf{only some} of the approaches.}
    \label{result:quant_results}
    \centering
    \begin{tabular}{llll}
        \toprule
                                
            & $\mathcal{L}_{\text{ccc}}(m)$    & $\mathcal{L}_{\text{ccc}}(s)$    & $\mathcal{L}_\text{KL}$     \\
        \midrule
        
        STL  \cite{han2020exploring}    & 0.719           & -                 & -               \\ 
        MTL PU  \cite{han2020exploring} & 0.734          & 0.286         & 0.797         \\ 
        MU                              & $\textbf{0.756}^{*}$ & 0.076          & 0.690         \\
        $+$LU                          & 0.744        & $\textbf{0.340}^{**}$    & $\textbf{0.258}^{**}$\\
        \bottomrule
    \end{tabular}
\end{table}

        
        

\subsection{Validation metrics}
To validate our mean estimates, we use $\mathcal{L}_\text{CCC}(m)$, a widely used metric in literature \cite{Tzirakis2018-speech, tzirakis2021-semspeech, han2020exploring}. To validate the label uncertainty estimates, we use $\mathcal{L}_\text{CCC}(s)$, along with $\mathcal{L}_\text{KL}$. $\mathcal{L}_\text{CCC}(s)$, previously used in \cite{han2017hard}, only validates the performance on $s$, and ignores performances on $m$. In this light, we additionally use $\mathcal{L}_\text{KL}$ which jointly validates both $m$ and $s$ performances. However, the metric makes a Gaussian assumption on $\mathcal{Y}_t$ and hence can be biased. In this light, we use both these metrics for the validation with their respective benefits under consideration. Statistical significance is estimated using one-tailed $t$-test, asserting significance for \emph{p}-values $\leq0.05$, similar to \cite{sridhar2021generative}.

\begin{figure}[t!]
    \centering
    \includegraphics[width=0.48\textwidth]{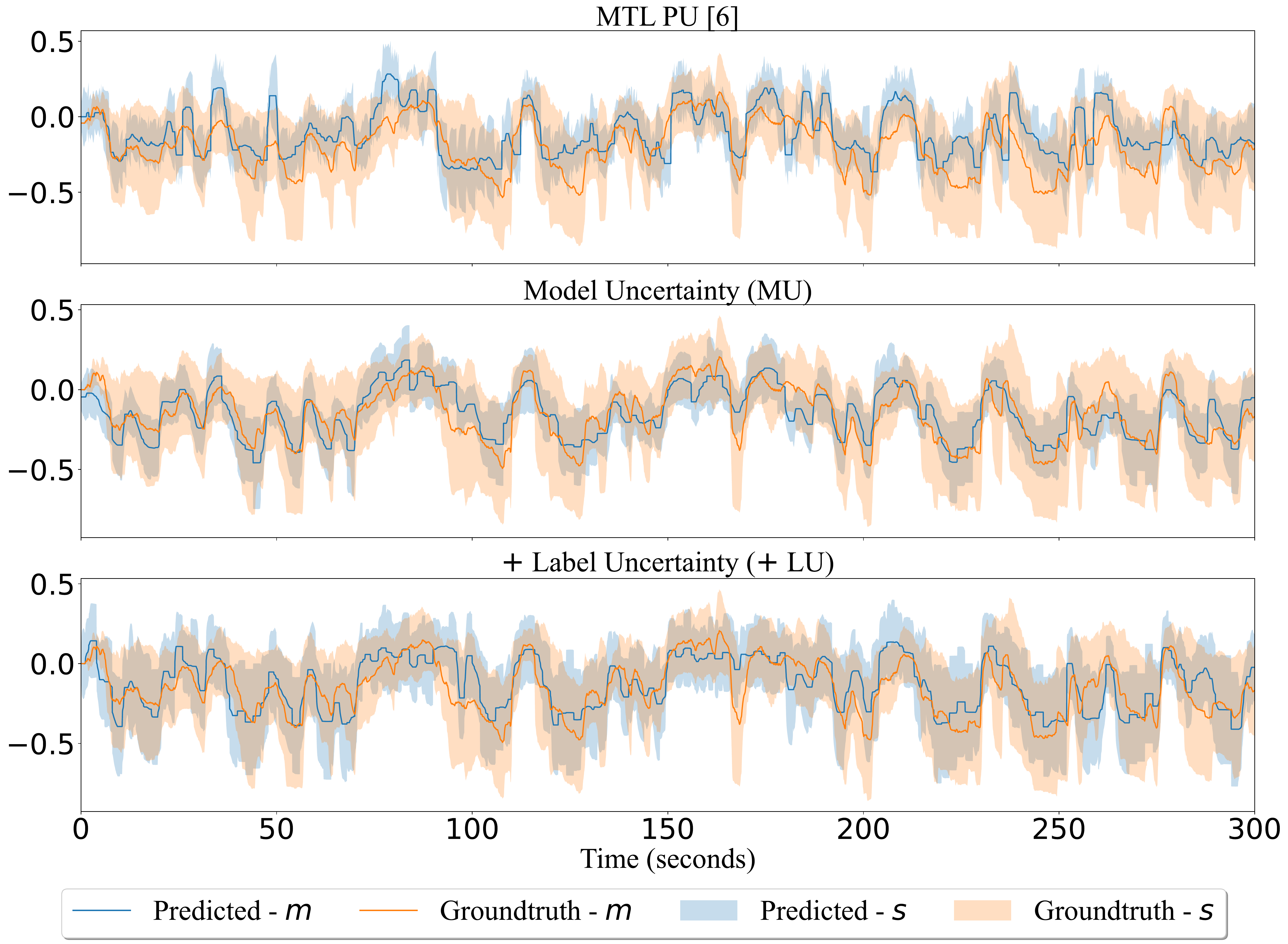}
    \caption{Results obtained for a test subject for arousal.}
    \captionsetup{justification=centering}
    \label{Fig:results-pred-arousal}
\end{figure}



 
 

\section{Results and Discussion}
\label{sec:Discussion}



Table \ref{result:quant_results} shows the average performance of the baselines and the proposed models, in terms of their mean $m$, standard deviation $s$, and distribution $\widehat{\mathcal{Y}}_t$ estimations, $\mathcal{L}_{\text{CCC}}(m)$, $\mathcal{L}_{\text{CCC}}(s)$ and $\mathcal{L}_{KL}$ respectively. Comparisons with respect to $\mathcal{L}_{\text{CCC}}(s)$ and $\mathcal{L}_{KL}$ are not presented for the STL baseline as this algorithm does not contain uncertainty modeling and does not estimate $s$.

\subsection{Comparison with baselines}
From Table \ref{result:quant_results}, we observe that both the proposed uncertainty models, MU and $+$LU, provide improved mean $m$ estimations over the baselines, in-terms of $\mathcal{L}_{\text{CCC}}(m)$, with statistical significance. Crucially, we observe that our models provide better $\mathcal{L}_{\text{CCC}}(m)$ even in comparison with the STL baseline \cite{han2020exploring} which is not an uncertainty model and only estimates $m$ without accounting for the label uncertainty. While uncertainty models MU and $+$LU achieve $0.756$ and $0.744$ $\mathcal{L}_{\text{CCC}}(m)$ respectively, STL and MTL PU \cite{han2020exploring} achieve $0.734$ and $0.719$ respectively. 

Secondly, we note that the proposed label uncertainty model $+$LU achieves state-of-the-art results for standard deviation $s$ and distribution $\widehat{\mathcal{Y}}_t$ estimations, with statistical significance, achieving $0.340$ $\mathcal{L}_{\text{CCC}}(s)$ and $0.258$ $\mathcal{L}_{KL}$ respectively. 
The $+$LU model is trained on a more informative distribution of annotations $\mathcal{Y}_t$, in contrast to training on the $s$ estimate \cite{han2020exploring}, thereby leading to better capturing label uncertainty in arousal annotations.
This explains the capability of distribution learning, using the $\mathcal{L}_\text{KL}$ loss term, for modeling label uncertainty in SER. Also noting here that, in contrast to the baselines, the proposed model learns uncertainty dependent representations in an end-to-end manner for improved uncertainty estimates, inline with literature \cite{alisamir2021evolution} which recommends end-to-end learning for uncertainty modeling in SER. Conclusively, the results in-terms of $\mathcal{L}_{\text{CCC}}(m)$, $\mathcal{L}_{\text{CCC}}(s)$, and $\mathcal{L}_{KL}$ reveal the ability of the proposed $+$LU model to best capture label uncertainty, thereby also improving mean estimations in comparison to the baselines.

To further validate the results, we plot the distribution estimations for a test subject, seen in Figure \ref{Fig:results-pred-arousal}. From the figure, as the quantitative results suggest, we see that the MU and $+$LU models are superior to the baseline MTL PU in-terms of the distribution estimations. Specifically, the $+$LU captures the whole distribution of arousal annotations better than the MTL PU model, by well capturing the time-varying uncertainty without noisy estimates. This further highlights the advantage of training on $\mathcal{L}_\text{KL}$ loss term.

\subsection{Comparison between \emph{MU} and $+$\emph{LU}}


Between the proposed uncertainty models, we note that $+$LU outperforms MU in terms of $\mathcal{L}_{\text{CCC}}(s)$, without significant degradations to $\mathcal{L}_{\text{CCC}}(m)$. Specifically, $+$LU improves drastically and significantly in terms of $\mathcal{L}_{\text{CCC}}(s)$ from $0.076$ to $0.340$. At the same time, for $\mathcal{L}_{\text{CCC}}(m)$ a slight reduction from $0.756$ to $0.744$ can be observed which, however, is not statistically significant. This reveals that, by also optimizing $\mathcal{L}_\text{KL}$, our model can better account for subjectivity in arousal. 
It is important to note here that the trade-off between $\mathcal{L}_{\text{CCC}}(s)$ and $\mathcal{L}_{\text{CCC}}(s)$ can be further adjusted using a different prior $P(w)$ on the weight distributions, as noted by Blundell et al. \cite{blundell2015weight} who recommend a spike-and-slab for mean centered predictions. However, in our case, as the arousal annotations do not follow a mean centered distribution (seen in Section \ref{sec:dataset}), we chose a simple Gaussian prior with unit standard deviation $\mathcal{N}(0, 1)$ to better capture the annotation distribution. Moreover, the choice of such as simple prior makes our model also scalable, eliminating the requirement to tune the prior with respect to different SER datasets.


Moreover, from the Figure \ref{Fig:results-pred-arousal}, backing the results in Table \ref{result:quant_results}, one may see that $+$LU, explicitly trained on $\mathcal{L}_\text{KL}$, best captures the subjectivity in emotions, while MU is optimized more for predictions centered on the mean and strict standard deviations.

\subsection{Label Uncertainty BNN for SER}


This work is the first in literature to study BNNs for SER.
The BBB technique, adopted by the proposed model, use simple gradient updates and produce stochastic outputs, making them promising candidates for end-to-end uncertainty modeling. Crucially, they open up possibilities for training the model on a distribution of annotations, rather than a less informative standard deviation estimate. Moreover, unlike the MTL PU which requires dataset-dependent tuning of the loss function, using the correlation estimate between $m$ and $s$ \cite{han2020exploring}, our proposed models do not require loss function tuning and are scalable across SER datasets with a simple prior initialization.

While the proposed model has several advantages, it also leaves room for future work. Firstly, the heuristics-based initialization of $P(w|D)$ for optimal performances could be further studied. Secondly, in the model we assumed that $\mathcal{Y}_t$ follows a Gaussian distribution, and had only $a=6$ annotations available to model the distribution. This might sometimes lead to unstable $\mathcal{L}_\text{KL}$ during approximation of $\mathcal{Y}_t$, thereby affecting the training processes. Acknowledging that gaining more annotation is resource inefficient, as future work, we will investigate techniques to model stable $\mathcal{Y}_t$ with limited annotations.



\section{Conclusions}
\label{sec:conclusion}

We introduced a BNN-based end-to-end approach for SER, which can account for the subjectivity and label uncertainty in emotional expressions. To this end, we introduced a loss term based on the KL divergence to enable our approach to be trained on a distribution of annotations. Unlike previous approaches, the stochastic outputs of our approach can be employed to estimate statistical moments such as the mean and standard deviation of the emotion annotations. Analysis of the results reveals that the proposed uncertainty model trained on the KL loss term can aptly capture the distribution of arousal annotations, achieving state-of-the-art results in mean and standard deviation estimations, in-terms of both the CCC and KL divergence metrics. 

\section{Acknowledgements}
The authors thank the Landesforschungsförderung Hamburg (LFF-FV79) for supporting this work under the ''Mechanisms of Change in Dynamic Social Interaction'' project.




\bibliographystyle{IEEEtran}
\bibliography{mybib}

\begin{thebibliography}{10}
\providecommand{\url}[1]{#1}
\csname url@samestyle\endcsname
\providecommand{\newblock}{\relax}
\providecommand{\bibinfo}[2]{#2}
\providecommand{\BIBentrySTDinterwordspacing}{\spaceskip=0pt\relax}
\providecommand{\BIBentryALTinterwordstretchfactor}{4}
\providecommand{\BIBentryALTinterwordspacing}{\spaceskip=\fontdimen2\font plus
\BIBentryALTinterwordstretchfactor\fontdimen3\font minus
  \fontdimen4\font\relax}
\providecommand{\BIBforeignlanguage}[2]{{%
\expandafter\ifx\csname l@#1\endcsname\relax
\typeout{** WARNING: IEEEtran.bst: No hyphenation pattern has been}%
\typeout{** loaded for the language `#1'. Using the pattern for}%
\typeout{** the default language instead.}%
\else
\language=\csname l@#1\endcsname
\fi
#2}}
\providecommand{\BIBdecl}{\relax}
\BIBdecl

\bibitem{ledoux2018subjective}
J.~E. LeDoux and S.~G. Hofmann, ``The subjective experience of emotion: a
  fearful view,'' \emph{Current Opinion in Behavioral Sciences}, vol.~19, pp.
  67--72, 2018.

\bibitem{lei2015affect}
Z.~Lei and N.~Lehmann-Willenbrock, ``Affect in meetings: An interpersonal
  construct in dynamic interaction processes,'' in \emph{The Cambridge handbook
  of meeting science}.\hskip 1em plus 0.5em minus 0.4em\relax Cambridge
  University Press, 2015, pp. 456--482.

\bibitem{nummenmaa2018maps}
L.~Nummenmaa, R.~Hari, J.~K. Hietanen, and E.~Glerean, ``Maps of subjective
  feelings,'' \emph{Proceedings of the National Academy of Sciences}, vol. 115,
  no.~37, pp. 9198--9203, 2018.

\bibitem{Schuller2018-xi}
B.~W. Schuller, ``{Speech emotion recognition: two decades in a nutshell,
  benchmarks, and ongoing trends},'' \emph{Communications of the ACM}, vol.~61,
  no.~5, pp. 90--99, Apr. 2018.

\bibitem{dukes2021rise}
D.~Dukes, K.~Abrams, R.~Adolphs, M.~E. Ahmed, A.~Beatty, K.~C. Berridge,
  S.~Broomhall, T.~Brosch, J.~J. Campos, Z.~Clay \emph{et~al.}, ``The rise of
  affectivism,'' \emph{Nature Human Behaviour}, pp. 1--5, 2021.

\bibitem{han2020exploring}
J.~Han, Z.~Zhang, Z.~Ren, and B.~Schuller, ``{Exploring perception uncertainty
  for emotion recognition in dyadic conversation and music listening},''
  \emph{Cognitive Computation}, pp. 1--10, 2020.

\bibitem{sridhar2020modeling}
K.~Sridhar and C.~Busso, ``{Modeling uncertainty in predicting emotional
  attributes from spontaneous speech},'' in \emph{ICASSP - IEEE Int. Conf. on
  Acoustics, Speech and Signal Processing}, 2020, pp. 8384--8388.

\bibitem{Tzirakis2018-speech}
P.~Tzirakis, J.~Zhang, and B.~W. Schuller, ``{End-to-End Speech Emotion
  Recognition Using Deep Neural Networks},'' in \emph{ICASSP - Int. Conf. on
  Acoustics, Speech and Signal Processing}, 2018, pp. 5089--5093.

\bibitem{tzirakis2021-mm}
P.~Tzirakis, J.~Chen, S.~Zafeiriou, and B.~Schuller, ``{End-to-end multimodal
  affect recognition in real-world environments},'' \emph{Information Fusion},
  vol.~68, pp. 46--53, 2021.

\bibitem{han2017hard}
J.~Han, Z.~Zhang, M.~Schmitt, M.~Pantic, and B.~Schuller, ``{From hard to soft:
  Towards more human-like emotion recognition by modelling the perception
  uncertainty},'' in \emph{Proceedings of the 25th ACM Int. Conf. on
  Multimedia}, 2017, pp. 890--897.

\bibitem{gunes2013categorical}
H.~Gunes and B.~Schuller, ``Categorical and dimensional affect analysis in
  continuous input: Current trends and future directions,'' \emph{Image and
  Vision Computing}, vol.~31, no.~2, pp. 120--136, 2013.

\bibitem{alisamir2021evolution}
S.~Alisamir and F.~Ringeval, ``On the evolution of speech representations for
  affective computing: A brief history and critical overview,'' \emph{IEEE
  Signal Processing Magazine}, vol.~38, no.~6, pp. 12--21, 2021.

\bibitem{zheng2021uncertainty}
R.~Zheng, S.~Zhang, L.~Liu, Y.~Luo, and M.~Sun, ``{Uncertainty in bayesian deep
  label distribution learning},'' \emph{Applied Soft Computing}, 2021.

\bibitem{kohl2018probabilistic}
S.~Kohl, B.~Romera-Paredes, C.~Meyer, J.~De~Fauw, J.~R. Ledsam, K.~Maier-Hein,
  S.~Eslami, D.~Jimenez~Rezende, and O.~Ronneberger, ``A probabilistic u-net
  for segmentation of ambiguous images,'' \emph{Advances in Neural Information
  Processing Systems}, vol.~31, 2018.

\bibitem{garnelo2018conditional}
M.~Garnelo, D.~Rosenbaum, C.~Maddison, T.~Ramalho, D.~Saxton, M.~Shanahan,
  Y.~W. Teh, D.~Rezende, and S.~A. Eslami, ``Conditional neural processes,'' in
  \emph{Int. Conf. on Machine Learning}.\hskip 1em plus 0.5em minus 0.4em\relax
  PMLR, 2018, pp. 1704--1713.

\bibitem{blundell2015weight}
C.~Blundell, J.~Cornebise, K.~Kavukcuoglu, and D.~Wierstra, ``{Weight
  uncertainty in neural network},'' in \emph{International Conference on
  Machine Learning}.\hskip 1em plus 0.5em minus 0.4em\relax PMLR, 2015, pp.
  1613--1622.

\bibitem{reisenzein1994pleasure}
R.~Reisenzein, ``Pleasure-arousal theory and the intensity of emotions.''
  \emph{Journal of personality and social psychology}, vol.~67, no.~3, p. 525,
  1994.

\bibitem{russell1980circumplex}
J.~A. Russell, ``A circumplex model of affect.'' \emph{Journal of personality
  and social psychology}, vol.~39, no.~6, p. 1161, 1980.

\bibitem{t21_interspeech}
M.~K. T, E.~Sanchez, G.~Tzimiropoulos, T.~Giesbrecht, and M.~Valstar,
  ``{Stochastic Process Regression for Cross-Cultural Speech Emotion
  Recognition},'' in \emph{Proc. Interspeech 2021}, 2021, pp. 3390--3394.

\bibitem{recolaDB}
F.~Ringeval, A.~Sonderegger, J.~Sauer, and D.~Lalanne, ``{Introducing the
  RECOLA multimodal corpus of remote collaborative and affective
  interactions},'' in \emph{10th IEEE Int. Conf. and Workshops on Automatic
  Face and Gesture Recognition (FG)}, 2013, pp. 1--8.

\bibitem{raj2020defining}
N.~Raj~Prabhu, C.~Raman, and H.~Hung, ``{Defining and Quantifying Conversation
  Quality in Spontaneous Interactions},'' in \emph{Comp. Pub. of 2020 Int.
  Conf. on Multimodal Interaction}, 2020, pp. 196--205.

\bibitem{abdelwahab2019active}
M.~Abdelwahab and C.~Busso, ``{Active learning for speech emotion recognition
  using deep neural network},'' in \emph{2019 8th Int. Conf. on Affective
  Computing and Intelligent Interaction (ACII)}.\hskip 1em plus 0.5em minus
  0.4em\relax IEEE, 2019.

\bibitem{avec16}
\BIBentryALTinterwordspacing
M.~Valstar, J.~Gratch, B.~Schuller, F.~Ringeval, D.~Lalanne, M.~Torres~Torres,
  S.~Scherer, G.~Stratou, R.~Cowie, and M.~Pantic, ``{AVEC 2016: Depression,
  Mood, and Emotion Recognition Workshop and Challenge},'' in \emph{Proceedings
  of the 6th International Workshop on Audio/Visual Emotion Challenge}, ser.
  AVEC '16.\hskip 1em plus 0.5em minus 0.4em\relax New York, NY, USA:
  Association for Computing Machinery, 2016, pp. 3--10. [Online]. Available:
  \url{https://doi.org/10.1145/2988257.2988258}
\BIBentrySTDinterwordspacing

\bibitem{grimm2005evaluation}
M.~Grimm and K.~Kroschel, ``{Evaluation of natural emotions using self
  assessment manikins},'' in \emph{IEEE Workshop on Automatic Speech
  Recognition and Understanding, 2005.}\hskip 1em plus 0.5em minus 0.4em\relax
  IEEE, 2005, pp. 381--385.

\bibitem{trigeorgis2016adieu}
G.~Trigeorgis, F.~Ringeval, R.~Brueckner, E.~Marchi, M.~A. Nicolaou,
  B.~Schuller, and S.~Zafeiriou, ``{Adieu features? end-to-end speech emotion
  recognition using a deep convolutional recurrent network},'' in \emph{2016
  Int. Conf. on acoustics, speech and signal processing (ICASSP)}.\hskip 1em
  plus 0.5em minus 0.4em\relax IEEE, 2016.

\bibitem{Goodfellow-et-al-2016}
I.~Goodfellow, Y.~Bengio, and A.~Courville, \emph{Deep Learning}, 3rd~ed., ser.
  5.\hskip 1em plus 0.5em minus 0.4em\relax The address of the publisher: MIT
  Press, 7 2016, vol.~4, ch.~3, pp. 51--77,
  \url{http://www.deeplearningbook.org}.

\bibitem{tzirakis2021-semspeech}
P.~Tzirakis, A.~Nguyen, S.~Zafeiriou, and B.~W. Schuller, ``Speech emotion
  recognition using semantic information,'' in \emph{ICASSP 2021-2021 IEEE
  International Conference on Acoustics, Speech and Signal Processing
  (ICASSP)}.\hskip 1em plus 0.5em minus 0.4em\relax IEEE, 2021, pp. 6279--6283.

\bibitem{sridhar2021generative}
K.~Sridhar, W.-C. Lin, and C.~Busso, ``Generative approach using soft-labels to
  learn uncertainty in predicting emotional attributes,'' in \emph{2021 9th
  International Conference on Affective Computing and Intelligent Interaction
  (ACII)}.\hskip 1em plus 0.5em minus 0.4em\relax IEEE, 2021, pp. 1--8.

\end{thebibliography}

\end{document}